\documentclass[draftclsnofoot, onecolumn, comsoc, 12pt]{IEEEtran}
\IEEEoverridecommandlockouts
\usepackage{comment}
\usepackage[export]{adjustbox}
\usepackage{xurl}  
\usepackage[group-separator={,},group-minimum-digits=4]{siunitx}
\usepackage{lipsum}
\usepackage{algorithm}
\usepackage{color, colortbl}
\definecolor{Gray}{gray}{0.9}
\definecolor{White}{gray}{1}
\usepackage{makecell} 
\usepackage{array}
\newcolumntype{P}[1]{>{\centering\arraybackslash}m{#1}}
\newcolumntype{A}{>{\columncolor{White}}P}
\usepackage{algpseudocode}

\usepackage{cite}
\usepackage{kotex}
\usepackage{subfigure}
\usepackage{multirow}
\usepackage{amsmath,amssymb,amsfonts}
\usepackage{float}
\usepackage{mathtools}

\usepackage{amsthm}
\usepackage{graphicx}
\usepackage{textcomp}
\usepackage[utf8]{inputenc}
\usepackage{hyperref}
\usepackage[english]{babel}
\usepackage{booktabs}
\usepackage[noabbrev]{cleveref}
\def\BibTeX{{\rm B\kern-.05em{\sc i\kern-.025em b}\kern-.08em
    T\kern-.1667em\lower.7ex\hbox{E}\kern-.125emX}}

\def\tcb{\textcolor{black}}

\usepackage{caption} 
\usepackage{tabularx}

\title{Rainbow Beamforming for Wideband LEO Satellite Communications: Principles, Applications, and Technical Challenges}
\author{Juha Park, Hyungseok Ko, Haejung Kim, Namyoon Lee, \\Ian P. Roberts, H. Vincent Poor, and Wonjae Shin 
\thanks{
    }
    \thanks{J. Park, H. Ko, H. Kim, and W. Shin are with the School of Electrical Engineering, Korea University, Seoul 02841, South Korea 
    (email: {\texttt{\{juha, hsko99, haejung, wjshin\}@korea.ac.kr}}); N. Lee is with the Department of Electrical Engineering,  POSTECH, Pohang 37673, South Korea (e-mail: \texttt{nylee@postech.ac.kr}); I. P. Roberts is with the Department of Electrical and Computer Engineering, UCLA, Los Angeles, CA 90095, USA (e-mail: \texttt{ianroberts@ucla.edu});  H. V. Poor is with the Department of Electrical and Computer Engineering, Princeton University, Princeton, NJ 08544, USA (email: {\texttt{poor@princeton.edu}}). 
    }}

\begin{document} 
\markboth{}
{Shell \MakeLowercase{\textit{et al.}}: Bare Demo of IEEEtran.cls for IEEE Journals}
\maketitle
\begin{abstract}
Low Earth Orbit (LEO) satellite communications (SATCOM) has emerged as a key enabler of global connectivity for 6G networks. To overcome the significant path loss of space-to-ground links, high-gain directional beamforming (BF) is indispensable. As LEO systems evolve toward  wider bandwidths to support data-intensive applications, however, they encounter a fundamental physical limitation known as the beam-squint effect, which induces frequency-dependent beam misalignment. Conventionally, the beam-squint effect has been treated as a critical performance impairment that must be mitigated. This article {introduces} a paradigm shift in wideband LEO satellite systems by redefining beam-squint as a valuable source of \textit{frequency-spatial diversity} and {presents} the principles of \textit{rainbow BF}. Rather than {mitigating} beam squint, rainbow BF deliberately exploits it to generate frequency-dependent beams, enabling different frequency components to illuminate distinct spatial directions using only a single or a small number of radio frequency chains. 
By supporting dynamic frequency-spatial beam allocation, rainbow BF offers enhanced flexibility and scalability for wideband LEO SATCOM. 
We further illustrate the benefits of rainbow BF through three representative LEO SATCOM applications: i) massive multiple access to overcome the latency and throughput bottlenecks of conventional beam hopping; ii) integrated sensing and communications for simultaneous target detection and data transmission; and iii) rapid satellite acquisition to reduce search overhead and improve link reliability. Finally, we discuss key implementation challenges and outline promising future research directions for rainbow BF in wideband LEO SATCOM.
\end{abstract}

\section{Introduction}

\begin{figure}[!b]
\centering
\includegraphics[width=0.95\linewidth]{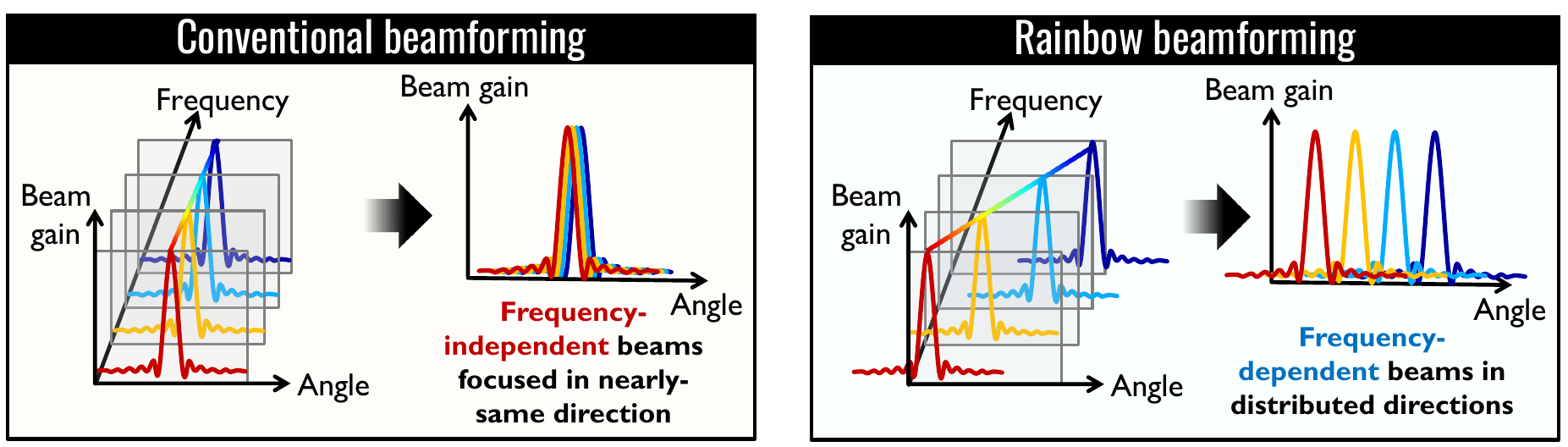}
\caption{Comparison of conventional BF using phased array based on PS with frequency-independent beams and rainbow BF with frequency-dependent beams across distributed directions.}
\label{fig1}
\end{figure}

\tcb{The growing demand for seamless global connectivity in beyond 5G (B5G) and 6G networks has positioned low Earth orbit (LEO) satellite communications (SATCOM) as a cornerstone technology  \cite{kodheli2020satellite}.} Operating at altitudes between \num{300} and \num{2000} km, LEO satellites offer significant advantages over traditional geostationary Earth orbit (GEO) and medium Earth orbit (MEO) systems, including reduced latency and enhanced link budgets due to shorter propagation distances.

{Although LEO satellites operate at lower altitudes than GEO or MEO systems, they still experience significant signal attenuation. For instance, at a \num{14} GHz carrier frequency with a satellite-user distance of \num{500} km, the free-space path loss reaches \num{169} dB. Therefore, high-gain beamforming (BF) is a fundamental requirement for LEO SATCOM and to achieve this under limited hardware resources, analog or analog-digital hybrid BF architectures have become essential.} These architectures provide high directional gain while requiring only a limited number of radio frequency (RF) chains. This makes analog or hybrid BF practical for LEO satellite constellations, where reducing the cost per satellite is essential for deploying tens of thousands of satellites.

\tcb{Another key challenge for LEO SATCOM is to match the data rates of commercial terrestrial networks such as 5G. Emerging applications, including ultra-high-definition video streaming, virtual/extended reality, and digital twins, demand very high data rates, requiring LEO systems to evolve toward wider bandwidths.} \tcb{This evolution to wideband LEO systems, aligned with the wideband targets of 6G non-terrestrial networks (NTN), however, introduces a fundamental technical challenge when combined with BF, namely the beam-squint effect.} In phased array architectures, different frequency components of a wideband signal are steered toward slightly different spatial directions. \tcb{This phenomenon arises because phase shifters provide \textit{frequency-independent} phase shifts, whereas the array response is inherently \textit{frequency-dependent}.} \tcb{In principle, fully digital BF can completely mitigate beam-squint through frequency-dependent precoding. However, it requires multiple RF chains, which leads to prohibitive power consumption and hardware complexity and makes it impractical for onboard satellite implementation.}

Conventional research on beam-squint has predominantly focused on compensation-oriented approaches that aim to eliminate or mitigate its detrimental effects. In the digital domain, frequency-dependent precoding techniques \cite{yu2021performance,chen2020hybrid}, such as per-subcarrier precoding, have been proposed to correct the frequency-dependent beam misalignment. In the analog domain, true time delay (TTD)-based architectures have been widely adopted to suppress beam-squint effects across wide bandwidths. For example, commercial SATCOM companies such as SatixFy employ TTD architectures to compensate for beam-squint in practical systems \cite{pozar2018multibeam}. More recently, however, a paradigm shift has emerged in which beam-squint is exploited as a valuable system resource rather than being treated as an impairment. This innovative approach, referred to as \emph{rainbow BF}, deliberately leverages beam-squint effects to realize frequency-dependent BF as illustrated in Fig. \ref{fig1}, with only a single or a few RF chains. To achieve this, joint phase-time array (JPTA) architectures that combine phase shifters {(PSs)} and {TTDs} have attracted significant attention and have been studied \cite{ratnam2022joint}.
By steering beams associated with different frequency components toward distinct spatial directions, rainbow BF {introduces} \textit{frequency-spatial diversity}, effectively decoupling frequency and spatial resources in beam allocation. This {is} in sharp contrast to conventional BF, where all frequency components are inherently coupled in the spatial domain and beams are confined to nearly identical directions, resulting in a significant lack of frequency-spatial diversity. \tcb{Exploiting this newly unlocked degree of freedom, several recent studies have demonstrated notable performance gains enabled by rainbow BF \cite{gao2023integrated,luo2024yolo,kim2023fast,10988651}.} 

\tcb{For integrated sensing and communications (ISAC), \cite{luo2024yolo} proposed a one-shot sensing method using rainbow beams in the terahertz band. For terrestrial mmWave networks, \cite{kim2023fast} proposed a rainbow BF-enabled fast beam management method that significantly reduces beam training latency compared to conventional hierarchical codebook-based approaches.}

\tcb{The majority of rainbow BF research has focused on terrestrial wireless networks. Despite its promising potential, rainbow BF techniques for LEO SATCOM remain relatively unexplored. The unique characteristics of LEO SATCOM present both compelling opportunities and fundamental challenges for the adoption of rainbow BF. On the one hand, LEO satellites have an inherently wide coverage footprint, substantially larger than that of terrestrial base stations. This makes rainbow BF-empowered multiple access particularly attractive \cite{park2025embracing, 10988651}.} \tcb{This advantage is further reinforced by the need to serve massive numbers of geographically distributed users under stringent onboard hardware constraints.} On the other hand, LEO SATCOM {introduces} system-specific phenomena that are largely absent in terrestrial networks. These include pronounced Doppler shifts \cite{park2025beyond} caused by high satellite velocity, extremely short channel coherence {times}, and the requirement for three-dimensional beam steering.

{In this article, we present a promising beamforming paradigm that transforms beam-squint from a long-standing impairment into a source of \textit{frequency-spatial diversity}, enhancing LEO SATCOM performance.} First, we introduce the fundamental principles of rainbow BF, including the design of frequency-direction mapping. \tcb{We then examine three promising applications in which rainbow BF offers attractive solutions: i) multiple access to address the latency and uplink throughput limitations of conventional beam hopping systems, ii) ISAC to enable joint radar and communication functions on a shared platform, and iii) satellite acquisition to reduce search overhead.} For each application scenario, we identify the associated technical challenges and discuss promising solutions. Furthermore, we discuss cross-cutting implementation issues, including cost-effective JPTA design, alternative architectures such as leaky wave antenna (LWA) \cite{jackson2012leaky}, {time-modulated} array (TMA) \cite{varma2023time}, and delay-adjustable intelligent reflecting surface (DA-IRS) \cite{10988651}. By presenting both the transformative potential and practical design considerations, this article aims to provide research directions for realizing and exploiting rainbow BF in next-generation LEO satellite networks.

\section{Fundamentals of Rainbow Beamforming}

\begin{figure*}[!ht]
\centering
\includegraphics[width=0.95\linewidth]{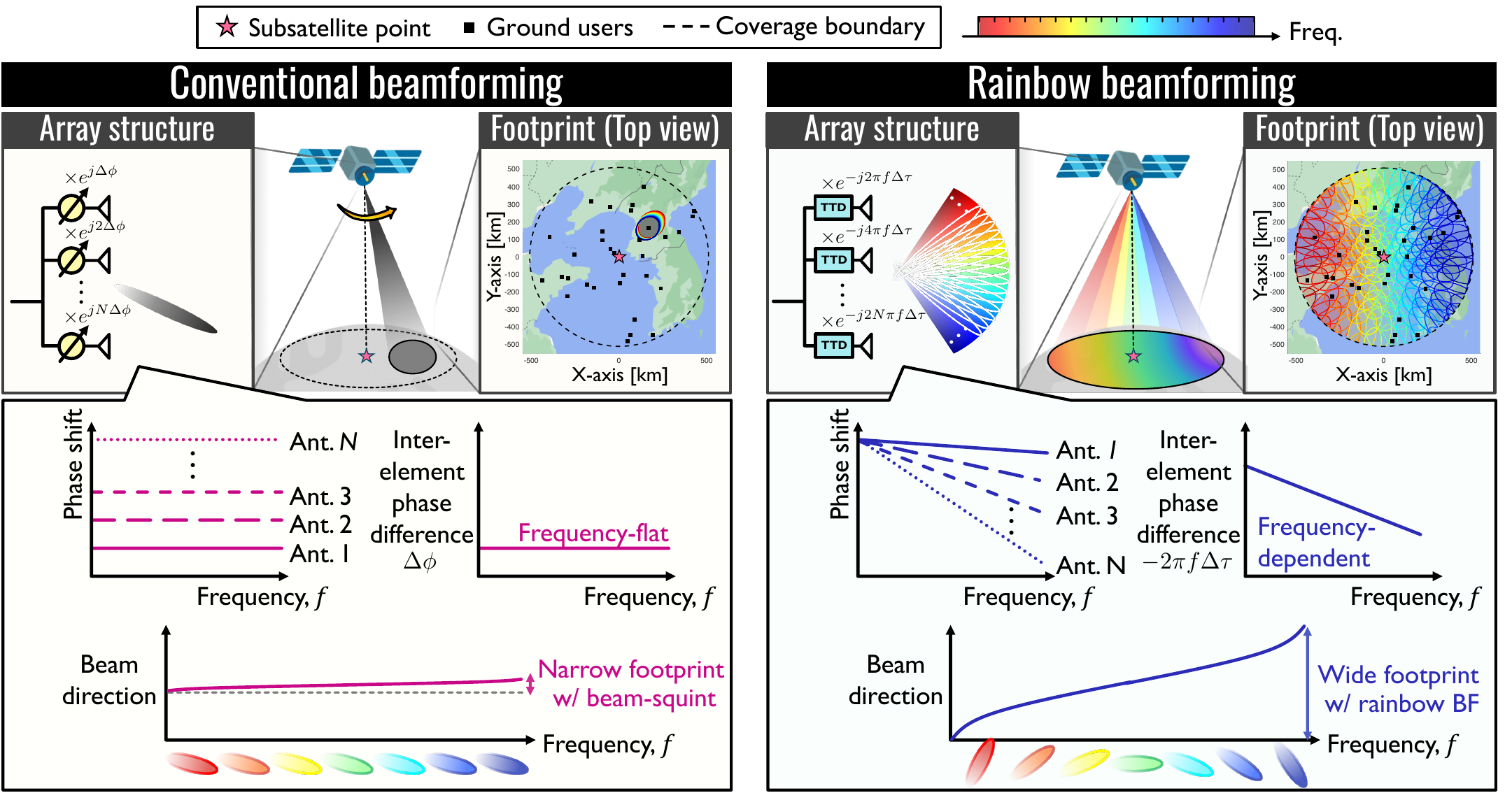}
\caption{\tcb{Comparison of conventional BF and rainbow BF: frequency-independent phase shifts with PSs versus frequency-dependent phase shifts with TTDs, and resulting beam footprints. TTD-only architecture is presented in rainbow BF for simplicity. The footprint plots are obtained for a uniform rectangular array (URA) with $8 \times 8$ antenna elements, a \num{14} GHz center frequency, \num{10} \% fractional bandwidth, and a satellite altitude of \num{500} km.}}
\label{fig2}
\end{figure*}
\subsection{{Principles of} Rainbow Beamforming}

\tcb{Rainbow BF is inspired by the optical phenomenon where different wavelengths of visible light refract at different angles, producing a spectrum of colors. Analogously, rainbow BF steers different frequency components of a wideband RF signal toward distinct spatial directions. The principle of BF is to compensate for the phase differences across antenna elements, thereby achieving constructive interference in a desired direction. Importantly, the resulting beam direction is determined by the relative phase differences between adjacent elements, not by the absolute phase shift applied at each element.}

Conventional BF based on PSs applies a frequency-flat phase shift to each antenna element. Specifically, when phase shifts of $\Delta \phi, 2\Delta \phi, \cdots, N\Delta\phi$ are applied across an $N$-element array, a constant inter-element phase difference of $\Delta \phi$ is maintained over the entire frequency band. As a result, all frequency components are steered toward nearly the same spatial direction, with only slight deviations caused by unwanted beam-squint effects, as illustrated in Fig. \ref{fig2}.

\begin{figure*}[!t]
\centering
\includegraphics[width=0.95\linewidth]{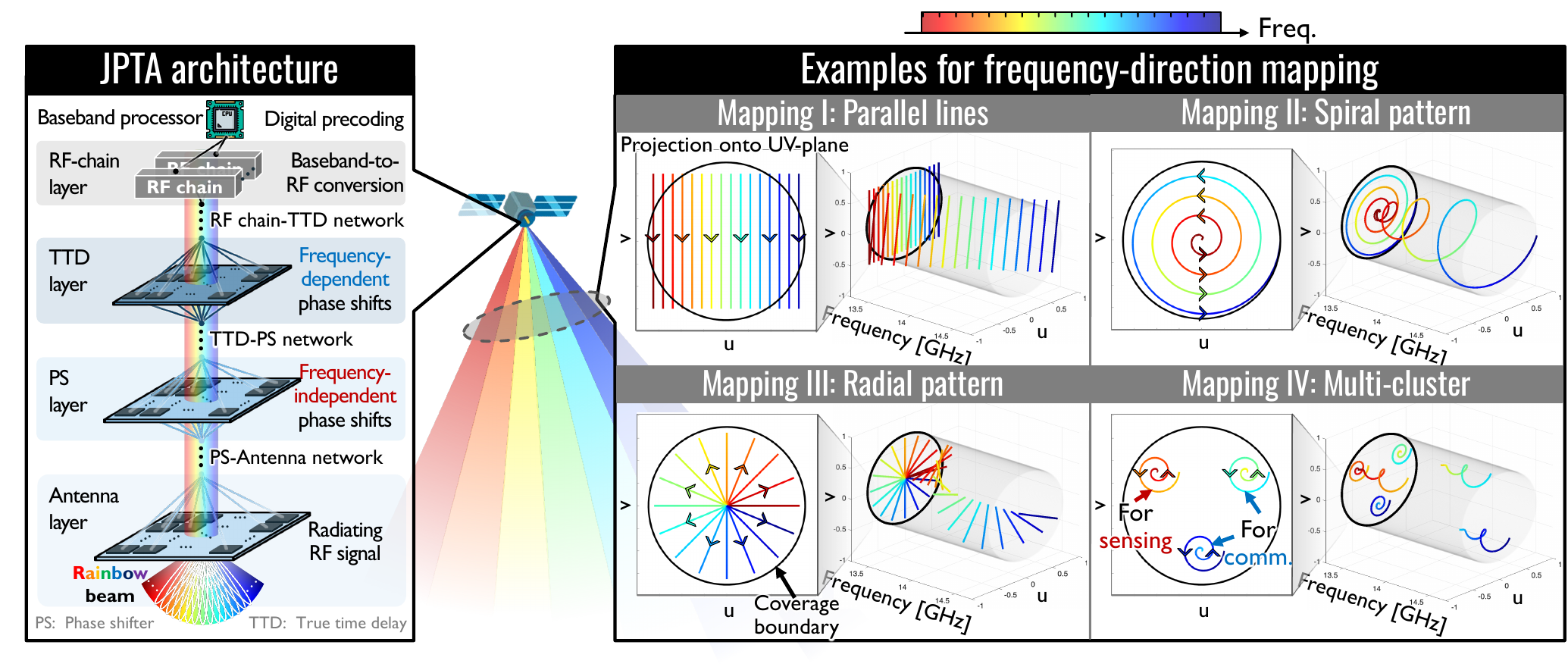}
\caption{\tcb{JPTA architecture and examples of frequency-direction mappings for rainbow BF. The deviation of the realized mapping from each desired mapping depends on system parameters such as the number of antennas and the array geometry.}}
\label{fig3}
\end{figure*}

In contrast, TTD elements introduce frequency-dependent inter-element phase differences. When progressive delays of $\Delta \tau, 2\Delta \tau, \cdots, N\Delta \tau$ are applied across the antenna array, the resulting phase shifts become $-2\pi f \Delta\tau, -4\pi f \Delta\tau, \cdots, -2N\pi f \Delta\tau,$ varying linearly with the frequency $f$. Consequently, the inter-element phase difference between adjacent antennas, given by $-2\pi f \Delta\tau$, also varies with frequency, where a larger $\Delta \tau$ yields a steeper slope of inter-element phase shift in the frequency domain. By properly selecting $\Delta \tau$, rainbow BF can be achieved, enabling beam directions to vary significantly across the operating bandwidth. \tcb{Note that this TTD-only architecture with a uniform linear array (ULA) is a simple example with progressively increasing delays; allowing independent delay values per element enables a more flexible design.} \tcb{Moreover, practical satellite systems require three-dimensional beam steering, which can also support the quasi Earth-fixed beam modes, necessitating array structures such as uniform planar arrays (UPAs).} In addition, JPTA architecture that combines both TTDs and PSs, as illustrated on the left side of Fig. \ref{fig3}, can provide even greater design flexibility for 
rainbow BF.

Fig. \ref{fig2} highlights the key advantage of rainbow BF. With conventional BF, the satellite's beam footprint on Earth's surface is narrow and concentrated, covering only a small region at any given time. To serve the entire coverage area, the satellite must perform sequential time-domain {beam switching (also called beam hopping).} In contrast, rainbow BF allows the satellite to simultaneously illuminate a wide area, where different frequency components form beam footprints at distinct spatial locations, collectively covering the entire region within a single time slot.

\begin{figure*}[!t]
\centering
\includegraphics[width=0.95\linewidth]{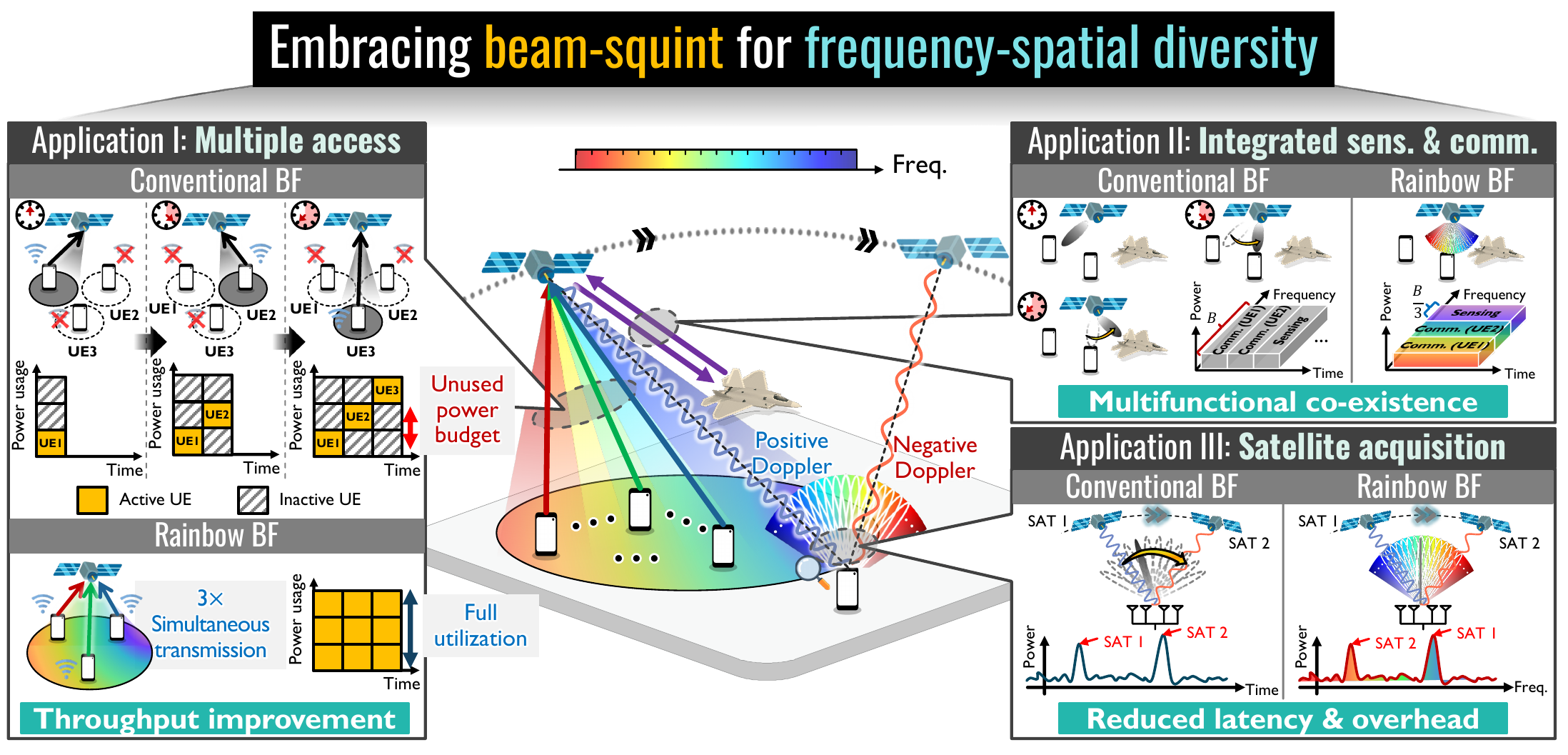}
\caption{Rainbow BF applications for LEO satellites: i) multiple access, ii) ISAC, and iii) satellite acquisition.}
\label{fig4}
\end{figure*}

\subsection{Frequency-Direction Mapping Design}
As discussed earlier, beam direction is determined by the relative phase differences between antenna elements, and JPTA architecture enables frequency-dependent control of these differences to implement rainbow BF. A central design question is therefore determining the beam direction associated with each frequency component, referred to as \textit{frequency-direction mapping}, which specifies the desired beam spatial direction as a function of frequency \cite{park2025embracing}. Designing efficient frequency-direction mappings is crucial for all rainbow BF applications. Representative examples of frequency-direction mappings are illustrated on the right side of Fig. \ref{fig3}. When the objective is to collectively cover an entire coverage area using beams formed by different frequency components, various spatial sweeping patterns, such as mappings I--III, can be employed. Alternatively, to serve multiple user clusters or to perform simultaneous communication and sensing, the frequency band can be partitioned into multiple sub-bands, each steered toward a different spatial direction, as shown in mapping IV. The choice of mapping has a direct impact on system performance and is application dependent: for multiple access, it affects coverage area and user capacity, whereas for sensing applications, it influences detection probability and  localization accuracy. In addition, 
practical constraints such as bandwidth and system requirements must be taken into account. 

From an implementation perspective,  developing algorithms that can realize the desired frequency-direction mapping with acceptable computational complexity for on-board processing remains challenging. In \cite{park2025embracing}, a JPTA-based analog rainbow BF algorithm was proposed that takes the frequency-direction mapping as input and outputs optimized TTD and PS values, achieving a computation time of approximately \num{0.06} seconds for \num{1024} subcarriers and \num{64} antenna elements. Nevertheless, the joint design of frequency-direction mapping and JPTA BF parameter optimization remains an important and open research direction.

\section{Promising Applications of Rainbow BF}
\tcb{Exploiting the frequency-spatial diversity of rainbow BF, this section examines three application scenarios, illustrated in Fig.~\ref{fig4}: i) multiple access, ii) ISAC, and iii) satellite acquisition.}

\begin{figure}[!t]
\centering
\includegraphics[width=0.95\linewidth]{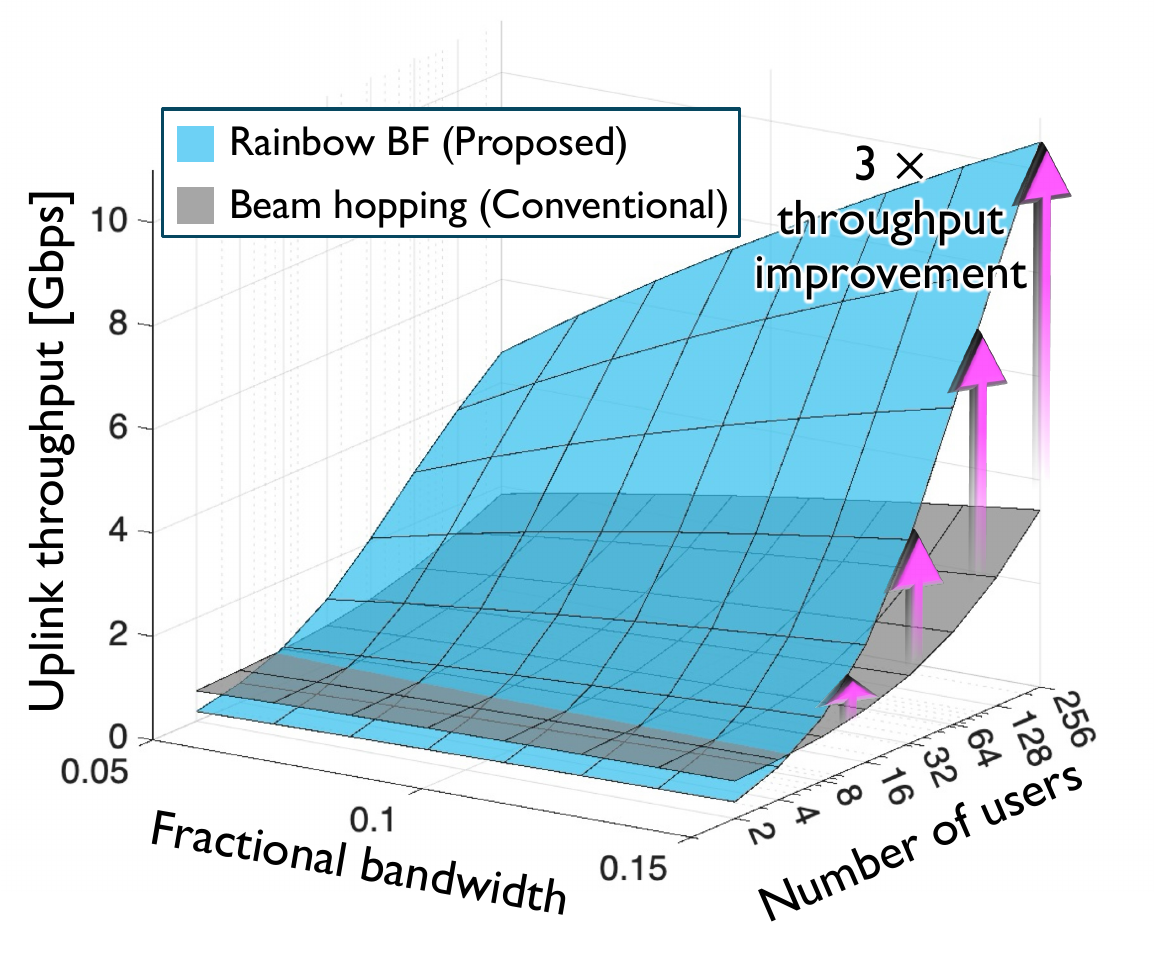}
\caption{\tcb{Performance comparison of rainbow BF and beam hopping for a LEO satellite with \num{64} antennas and single RF chain, using \num{1024} subcarriers at \num{14} GHz center frequency. Fractional bandwidth is defined as the ratio of bandwidth to center frequency. Mapping I in Fig. 3 is employed for rainbow BF optimization. Orthogonal frequency-division multiple access (OFDMA) is adopted, where each subcarrier is exclusively allocated to at most one user. Simulation parameters are the same as in \cite{park2025embracing}.}}
\label{fig5}
\end{figure}

\subsection{Application I: Massive Multiple Access}
To address limited onboard RF chains and stringent power budgets, conventional LEO satellite systems predominantly rely on beam hopping to serve a large number of users with a limited number of simultaneous active beams. Beam hopping employs a time-domain switching mechanism, where high-gain spot beams sequentially scan the service area. \tcb{This approach concentrates power on a single beam direction at each time instant. However, because it serves users in a time-division manner, it becomes a fundamental bottleneck for massive multiple access.} 

\tcb{Throughput bottleneck is particularly critical in uplink scenarios. Traditionally, satellites have mainly filled coverage holes of terrestrial networks while supporting low-rate services. However, the recent shift toward broadband and user-generated traffic may make the uplink an increasingly important bottleneck.} \tcb{In any given time slot, under beam hopping, only the users within the active beam transmit, while the remaining users stay idle. As a result, the idle users contribute no uplink power, which leads to high access latency and severe underutilization of the aggregate uplink power.}

\tcb{Rainbow BF fundamentally changes this operating paradigm, since its frequency-dependent beams point toward different directions and thus jointly cover a wide area. As a result, the satellite serves the entire coverage region within a single time slot. This allows spatially distributed users to access the network at the same time, each on its own assigned frequency band.} \tcb{As shown in Fig. \ref{fig5}, where each user transmits with its own power budget while the satellite performs reception, rainbow BF maximizes the aggregate uplink power budget by enabling concurrent uplink transmissions, thereby significantly boosting uplink throughput.} Notably, this gain becomes more pronounced as the number of users or the system bandwidth increases, highlighting the scalability of rainbow BF-empowered massive multiple access. \tcb{This characteristic aligns well with the requirements of LEO SATCOM systems, which are increasingly expected to support uplink-heavy traffic from a large number of users.} In addition to throughput improvements, rainbow BF substantially alleviates the latency constraints inherent to time-domain beam switching.

Despite these advantages, several challenges remain. A key issue is the optimization of frequency-direction mapping for multiple access. While Fig. \ref{fig3} illustrates simple mappings that do not account for user-specific information, practical deployments must consider heterogeneous user power budgets, geographical distributions, and traffic demands.  This necessitates the joint optimization of frequency-direction mapping, analog beamforming parameters (i.e., TTD and PS parameters), and resource allocation, leading to a highly coupled and computationally challenging optimization problem. Extending rainbow BF to scenarios with multiple RF chains also remains an open research direction. With two or more RF chains, additional degrees of freedom become available through digital-domain precoding or combining, which can be leveraged to reduce the gap between the desired frequency-direction mapping and the actually achieved mapping. Another promising direction is the design of frequency-direction mappings that allow specific frequency components to form beams toward multiple spatial directions. By enabling non-exclusively frequency bands among users, advanced multiple access techniques such as spatial-division multiple access (SDMA), non-orthogonal multiple access (NOMA), or rate-splitting multiple access (RSMA) can be employed in the digital domain to achieve further performance gains.

\subsection{Application II: Integrated Sensing and Communications}

ISAC is an emerging technology that combines sensing and communication functionalities using shared hardware and spectrum, thereby alleviating spectrum congestion  and improving hardware and resource efficiency \cite{gao2023integrated}. In LEO satellite systems, rainbow BF's frequency-spatial diversity offers a unique advantage for ISAC applications. {Specifically, when rainbow BF disperses downlink beams into distinct spatial directions according to frequency}, as illustrated in the center of Fig. \ref{fig4}, radar targets such as drones or aircraft reflect signals at frequency-dependent angles. By identifying the frequency components that produce strong reflections, the satellite can infer target locations using the known frequency-direction mapping. Since the process does not rely on time-domain sweeping, sensing information can be updated rapidly, enabling low-latency target detection and tracking. 

LEO satellite-based sensing inherently faces severe path loss challenges. In particular, monostatic sensing experiences round-trip path loss that scales with the fourth power of distance, making reliable target detection particularly challenging. To mitigate this limitation, bistatic ISAC has emerged as a promising solution by employing a ground-based receiver, which significantly reduces the propagation distance because targets typically operate at much lower altitudes than satellites \cite{park2025bistatic}. Rainbow BF naturally supports bistatic ISAC, where the satellite transmits frequency-dependent beams in different directions and the ground-based receiver estimates target angles directly from the echo signal frequency. This enables efficient sensing without beam switching while  substantially improving the received echo signal strength.

A critical challenge arises, {however}, when targets have significant mobility. Moving targets introduce Doppler shifts that alter the echo frequency, thereby distorting the frequency-direction mapping and causing localization ambiguity. Designing frequency-direction mappings that are robust to Doppler-induced uncertainty while maintaining communication performance remains an open research problem. Furthermore, adaptive frequency-direction mapping strategies that balance the communication-sensing trade-off require further investigation.

\begin{table*}[!t]
\caption{Comparison of Representative Implementation Architectures for Rainbow Beamforming}
\label{tab:rainbow_comparison}
\centering
\renewcommand{\arraystretch}{1.3}
\begin{tabularx}{\textwidth}{|>{\centering\arraybackslash}m{1.5cm}|>{\centering\arraybackslash}m{3.7cm}|>{\centering\arraybackslash}m{4.15cm}|>{\centering\arraybackslash}m{3cm}|>{\centering\arraybackslash}m{2cm}|}
\hline
\rowcolor[gray]{0.9}
\textbf{Architecture} & \textbf{Illustration} & \textbf{Operating Principle} & \textbf{Pros (+)} & \textbf{Cons (-)} \\ \hline
\textbf{Joint phase-time array (JPTA)} \cite{ratnam2022joint} & 
\includegraphics[width=3.2cm]{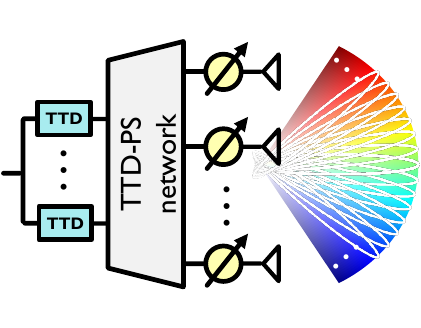} & 
Cascades TTDs and PSs to achieve flexible frequency-dependent beam patterns, where TTDs introduce phase shifts that vary linearly with frequency, while PSs provide a constant phase bias. & 
Flexible frequency-direction mapping and compatibility with conventional multiplexing (e.g., orthogonal frequency division multiplexing) & 
High hardware cost and power consumption of TTD \\ \hline
\textbf{Leaky wave antenna (LWA)} \cite{jackson2012leaky} & 
\includegraphics[width=3.2cm]{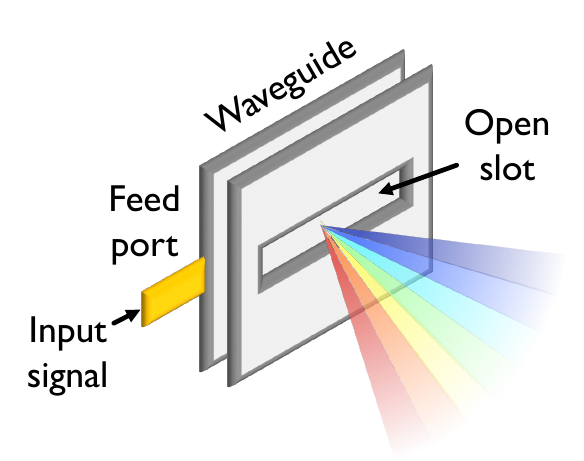} & 
Exploits the dispersive property of waveguide structures, where signal propagation characteristics vary with frequency, resulting in inherently frequency-dependent beam steering. & 
Passive structure without active RF components and low hardware cost & 
Fixed frequency-direction mapping \\ \hline
\textbf{Time-modulated array (TMA)} \cite{varma2023time} & 
\includegraphics[width=3.2cm]{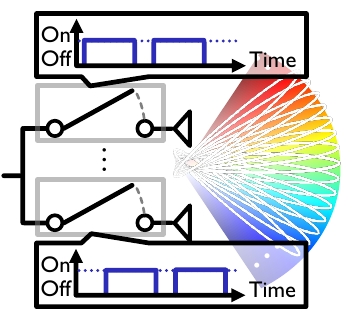} & 
Periodically switches antenna elements on/off, generating harmonics. Each harmonic beam is steered by controlling switching pattern. & 
Requires only switches (no TTD or PS)
& 
Limited to single-stream operation, as harmonic sidebands are inherently coupled and cannot carry independent data \\ \hline
\textbf{Delay-adjustable IRS (DA-IRS)} \cite{10988651} & 
\includegraphics[width=3.2cm]{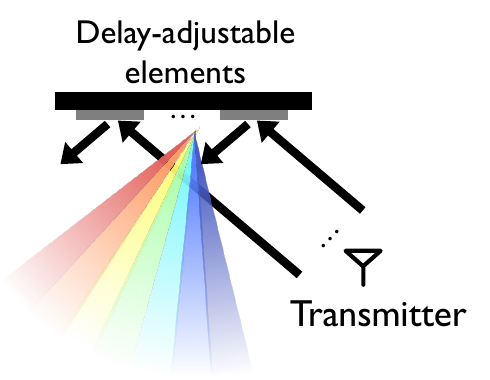} & 
Employs metamaterial-based structures with programmable delay elements to achieve phase shift and time delay control. Each reflective element can control both phase and delay of incident signals, enabling frequency-dependent reflection. & Lightweight and low (or near-zero) power consumption & High technical complexity in hardware implementation and difficulty in precise calibration
\\ \hline 
\end{tabularx}
\end{table*}

\subsection{Application III: Rapid Satellite Acquisition}

\tcb{Satellite acquisition is a critical process during initial access and handover, aiming to identify available satellites and estimate their angular positions for accurate beam steering at the ground terminal \cite{11105432}. Conventional approaches rely on time-domain sequential beam sweeping, where the receive beam  is sequentially steered across different directions to measure received power, and the satellite is detected in the direction corresponding to the maximum received power.} This sequential search, however, incurs substantial overhead and latency. {In} contrast, rainbow BF transforms this sequential process into a few-shot operation by simultaneously probing a wide angular region within a single time slot through frequency-spatial diversity.

In addition, this approach exploits Doppler effects, which are traditionally treated as impairments. In LEO SATCOM, the high velocity of satellites induces substantial Doppler shifts. Rather than compensating for these effects, rainbow BF leverages the quasi-deterministic relationship between a satellite's angular position and its Doppler shift, which can be predicted for a given orbit. Since different satellite positions correspond to distinct Doppler shifts, frequency-direction mappings can be designed such that beams at different frequencies align with the expected Doppler signatures, enabling reception from satellites over a wide angular range without beam sweeping. For a simplified two-dimensional geometry with a ULA, a closed-form rainbow beamformer was derived in \cite{park2025beyond}, where the frequency-direction mapping is explicitly matched to the Doppler-angle relationship. After receiving pilot signals, satellite angular positions are estimated using conventional frequency estimation methods combined with known frequency-direction mapping. \tcb{Numerical results in \cite{park2025beyond} show that rainbow BF can achieve higher estimation accuracy within a single time slot than conventional beam sweeping using \num{1024} time slots (assuming one beam switch per time slot), demonstrating a substantial reduction in acquisition overhead.}

The primary challenge lies in applying this approach to realistic three-dimensional geometry with satellites in multiple orbits. In such scenarios, Doppler shift alone does not uniquely determine satellite position, as satellites at different angular positions or orbital altitudes may produce identical Doppler shifts. To resolve this ambiguity, additional information such as the Doppler rate, which represents the time derivative of Doppler shift, can be exploited. Since Doppler rate depends on satellite trajectory and orbital parameters, it provides complementary information that {can be used to localize satellites}.
Nevertheless, developing a systematic framework and three-dimensional Doppler-aware frequency-direction mappings for multi-orbit scenarios remains an open research challenge.

\section{Implementation Issues and Open Challenges}

This section examines implementation architectures and cross-cutting challenges for the practical deployment of rainbow BF.

\subsection{Implementation Architectures}

\subsubsection{Cost-Effective JPTA Design}
As illustrated in Fig. \ref{fig3}, JPTA architectures offer multiple design options depending on the inter-connection structure among RF chains, TTDs, PSs, and antenna element layers. The design of cost-effective architectures is crucial, as TTDs are generally more expensive than PSs in terms of both power consumption and hardware costs \cite{park2025embracing}. To reduce implementation complexity, low-cost TTD hardware, such as fixed TTDs with non-adjustable delays or discrete TTDs supporting only a limited set of predefined delay values, can be considered. Since the feasible range and resolution of {a} TTD's delay values directly affect hardware costs, these constraints should be incorporated into frequency-direction mapping and beamformer optimization, which remains an important topic for future research.

\subsubsection{Alternatives to JPTA}
While JPTA offers flexible and reconfigurable frequency-direction mapping, alternative architectures can also generate frequency-dependent beams, as summarized in Table \ref{tab:rainbow_comparison}. LWA achieves frequency-dependent beam steering through waveguide dispersion without requiring active RF components \cite{jackson2012leaky}. TMA creates harmonic beams via periodic switching, eliminating the need for TTDs or PSs \cite{varma2023time}. In addition, DA-IRS leverages metamaterial-based surfaces with delay elements to decouple the beamforming aperture from the RF source, providing a lightweight and cost-effective solution \cite{10988651}. However, these approaches face limitations for LEO SATCOM: LWA lacks reconfigurability of frequency-direction mapping \cite{jackson2012leaky}, TMA cannot support multi-stream transmission in different directions \cite{varma2023time}, and DA-IRS presents high technical complexity in hardware implementation and difficulty in achieving precise wideband calibration \cite{10988651}.

\subsection{Cross-Application Open Issues}

\subsubsection{\tcb{Joint Frequency-Direction Mapping and Beamformer Optimization}}
\tcb{Most existing rainbow BF studies treat frequency-direction mapping design and JPTA beamformer optimization as separate problems. In practice,  how accurately a desired frequency-direction mapping can be achieved depends on hardware constraints such as maximum delay of TTD and the array structure, such as the array geometry, the number of antenna elements, and the antenna spacing. Joint optimization of frequency-direction mapping design and beamformer parameters can yield significant performance gains but leads to highly non-convex optimization problems with a large number of variables.}

\subsubsection{\tcb{Cross-Layer Resource Allocation for Rainbow Beamforming}} \tcb{Performance can be further improved by jointly optimizing rainbow BF with digital-domain processing, including subcarrier allocation, multiplexing, channel coding, power allocation, and user scheduling, although this comes at the cost of higher computational complexity. Since rainbow BF improves the effective channel quality in the analog domain, it can be readily combined with these digital-domain techniques and even create synergy through joint design.}

Emerging machine learning techniques offer promising approaches for such complex optimization. For example, deep reinforcement learning can learn effective frequency-direction mappings for given applications and scenarios. \tcb{An end-to-end framework that jointly optimizes frequency-direction mapping, analog/digital beamformers, advanced multiplexing, and resource allocation by considering user distributions, heterogeneous traffic demands, channel conditions, and orbital movements presents an interesting direction for future research.}

\subsubsection{Multi-Satellite Coordination}
Most current studies focus on single-satellite scenarios, whereas practical LEO systems consist of large constellations with overlapping coverage. Extending rainbow BF to multi-satellite operation introduces new challenges, particularly in managing inter-satellite interference. 
Coordinated frequency-direction mapping across neighboring satellites can mitigate such interference by assigning orthogonal or partially orthogonal frequency resources to overlapping coverage regions. Specifically, fractional frequency reuse allows non-overlapping frequency bands in overlapping areas while enabling full frequency reuse elsewhere,  thereby improving spectral efficiency and preventing coverage holes. Furthermore, rainbow BF naturally enables multi-satellite connectivity at the user terminal, allowing simultaneous communication with multiple satellites to achieve higher data rates and enhanced reliability.

\section{Conclusion}

Rainbow beamforming represents a paradigm shift in LEO SATCOM by transforming beam-squint from a long-lasting impairment into a valuable system resource. This article has outlined its fundamental principles and explored three promising applications: massive multiple access, ISAC, and rapid satellite acquisition. Although significant challenges remain in cost-effective JPTA design, joint optimization frameworks, and multi-satellite coordination, addressing these issues will be essential for enabling the practical deployment of rainbow BF in next-generation LEO satellite networks.


\bibliographystyle{IEEEtran}
\bibliography{bibfile}

\begin{thebibliography}{10}
\providecommand{\url}[1]{#1}
\csname url@samestyle\endcsname
\providecommand{\newblock}{\relax}
\providecommand{\bibinfo}[2]{#2}
\providecommand{\BIBentrySTDinterwordspacing}{\spaceskip=0pt\relax}
\providecommand{\BIBentryALTinterwordstretchfactor}{4}
\providecommand{\BIBentryALTinterwordspacing}{\spaceskip=\fontdimen2\font plus
\BIBentryALTinterwordstretchfactor\fontdimen3\font minus \fontdimen4\font\relax}
\providecommand{\BIBforeignlanguage}[2]{{%
\expandafter\ifx\csname l@#1\endcsname\relax
\typeout{** WARNING: IEEEtran.bst: No hyphenation pattern has been}%
\typeout{** loaded for the language `#1'. Using the pattern for}%
\typeout{** the default language instead.}%
\else
\language=\csname l@#1\endcsname
\fi
#2}}
\providecommand{\BIBdecl}{\relax}
\BIBdecl

\bibitem{kodheli2020satellite}
O.~Kodheli \emph{et~al.}, ``Satellite communications in the new space era: A survey and future challenges,'' \emph{IEEE Commun. Surveys Tuts.}, vol.~23, no.~1, pp. 70--109, 2021.

\bibitem{yu2021performance}
H.~Yu, P.~Guan, Y.~Wang, and Y.~Zhao, ``Performance analysis and codebook design for {mmWave} beamforming system with beam squint,'' \emph{IEEE Wireless Commun. Lett.}, vol.~10, no.~9, pp. 2013--2016, 2021.

\bibitem{chen2020hybrid}
Y.~Chen, Y.~Xiong, D.~Chen, T.~Jiang, S.~X. Ng, and L.~Hanzo, ``Hybrid precoding for wideband millimeter wave {MIMO} systems in the face of beam squint,'' \emph{IEEE Trans. Wireless Commun.}, vol.~20, no.~3, pp. 1847--1860, 2020.

\bibitem{pozar2018multibeam}
\BIBentryALTinterwordspacing
D.~Sikri and R.~M. Jayasuriya, ``Multi-beam phased array with full digital beamforming for {SATCOM} and {5G},'' \emph{Microwave Journal}, April 2019. [Online]. Available: \url{https://www.microwavejournal.com/articles/32053-multi-beam-phased-array-with-full-digital-beamforming-for-satcom-and-5g}
\BIBentrySTDinterwordspacing

\bibitem{ratnam2022joint}
V.~V. Ratnam \emph{et~al.}, ``Joint phase-time arrays: A paradigm for frequency-dependent analog beamforming in {6G},'' \emph{IEEE Access}, vol.~10, pp. 73\,364--73\,377, 2022.

\bibitem{gao2023integrated}
F.~Gao, L.~Xu, and S.~Ma, ``Integrated sensing and communications with joint beam-squint and beam-split for {mmWave}/{THz} massive {MIMO},'' \emph{IEEE Trans. Commun.}, vol.~71, no.~5, pp. 2963--2976, 2023.

\bibitem{luo2024yolo}
H.~Luo, F.~Gao, H.~Lin, S.~Ma, and H.~V. Poor, ``{YOLO}: An efficient terahertz band integrated sensing and communications scheme with beam squint,'' \emph{IEEE Trans. Wireless Commun.}, vol.~23, no.~8, pp. 9389--9403, 2024.

\bibitem{kim2023fast}
S.~Kim, J.~Park, J.~Moon, and B.~Shim, ``Fast and accurate terahertz beam management via frequency-dependent beamforming,'' \emph{IEEE Trans. Wireless Commun.}, vol.~23, no.~3, pp. 1699--1712, 2024.

\bibitem{10988651}
S.~Sekimori, H.~Hashida, Y.~Kawamoto, N.~Kato, K.~Yoshida, and M.~Ariyoshi, ``Deep {Q}-learning-driven frequency prism beamforming with delay-adjustable {IRS} in {LEO} satellite networks,'' \emph{IEEE Trans. Cogn. Commun. Netw.}, vol.~12, pp. 714--727, 2026.

\bibitem{park2025embracing}
J.~Park, S.~Kim, W.~Shin, and H.~V. Poor, ``Embracing beam-squint effects for wideband {LEO} satellite communications: A {3D} rainbow beamforming approach,'' \emph{IEEE Trans. Wireless Commun.}, vol.~25, pp. 16\,136--16\,152, 2026.

\bibitem{park2025beyond}
J.~Park, I.~P. Roberts, and W.~Shin, ``Beyond beam sweeping: One-shot satellite acquisition with {Doppler}-aware rainbow beamforming,'' \emph{IEEE Trans. Veh. Technol.}, pp. 1--6, 2026.

\bibitem{jackson2012leaky}
D.~R. Jackson, C.~Caloz, and T.~Itoh, ``Leaky-wave antennas,'' \emph{Proc. IEEE}, vol. 100, no.~7, pp. 2194--2206, 2012.

\bibitem{varma2023time}
D.~S. Varma, G.~Ram, and G.~A. Kumar, ``Time-modulated arrays: A review,'' \emph{IETE Tech. Rev.}, vol.~40, no.~1, pp. 136--151, 2023.

\bibitem{park2025bistatic}
J.~Park, J.~Seong, Y.~Mao, W.~Shin, and B.~Ottersten, ``A bistatic {ISAC} framework for {LEO} satellite systems: A rate-splitting approach,'' \emph{IEEE Trans. Aerosp. Electron. Syst.}, vol.~61, no.~6, pp. {17}{282}--{17}{301}, 2025.

\bibitem{11105432}
Y.~Zhang, X.~Ding, H.~Zhang, M.~Chen, and G.~Zhang, ``Initial access beam management framework for {LEO} satellite networks integrated with {5G} {NR},'' \emph{IEEE Internet of Things Journal}, vol.~13, no.~1, pp. 181--195, 2026.

\end{thebibliography}

\end{document}